# RISKS AND REMEDIES IN E-LEARNING SYSTEM

Nikhilesh Barik[1] and Dr. Sunil Karforma[2]

[1]Research Scholar, Department of Computer Science ,Burdwan University ,
West Bengal ,India
nikhileshbarik@gmail.com
[2] Reader and Head , Department of Computer Science ,
Burdwan University , West Bengal ,India
dr.sunilkarforma@gmail.com

## ABSTRACT

*One of the most effective applications of Information and Communication Technology (ICT) is the emergence of E-Learning. Considering the importance and need of E-Learning, recent years have seen a drastic change of learning methodologies in Higher Education. Undoubtedly ,the three main entities of E-Learning system can be considered as Student, Teacher & Controlling Authority and there will be different level, but a good E-Learning system needs total integrity among all entities in every level. Apart from integrity enforcement, security enforcement in the whole system is the other crucial way to organize the it. As internet is the backbone of the entire system which is inherently insecure , during transaction of message in E-Learning system, hackers attack by utilising different loopholes of technology. So different security measures are required to be imposed on the system. In this paper, emphasis is given on different risks called e-risks and their remedies called e-remedies to build trust in the minds of all participants of E-Learning system..*

## KEYWORDS

E-Learning, e-risks, e-remedies, E-Learning security,

## 1. INTRODUCTION

The application of **Information and Communication Technology (ICT)** in E-Learning environment has brought dramatic changes in higher education. In this context Internet meets the growing demand for advanced study material and associated resources. Any educational institution (University, Corporate Organization,etc) with some financial dealings is the matter of concern. But in the era of globalization, Students from different countries and communities may appear for the same degree or diploma as distance & geographic location is not a problem at all. So E-Learning is constructed in a variety of contexts, such as distance learning, online learning, and networked learning and learning to promote educational interactions between students, lecturers and learning communities[4].

E-Learning[14] means doing learning activities electronically through Internet. The development of a variety of E-Learning systems will change the higher education system entirely , especially with respect to the quality e-education services and support processes. In E-Learning system five significant participants are – Authors, Students, Managers, Teachers and System Developer (System Administrators). Hackers can change or modify the authenticated E-Learning documents like learning materials, certificates, question papers, lecture materials, mark sheets etc. which are communicated from Manager to Students and from Authors to Students as and when required. As technology has changed the current scenario of education system drastically, learners (in broader sense "Students") interested in education, are not confined to the conventional school, college and university campuses only.

A risk is the probability of the occurrence of a particular threat and the expected loss. E-risk involve the risk at the time of electronic transaction, whereas threat means an anticipated danger. Common threats for computers are viruses, network penetrations, theft and unauthorized modification of data, eavesdropping, and non-availability of servers and personal computers.

During transaction original documents may be modified ,tampered or destroyed by the hackers' active and passive attacks . Therefore knowledge of risk would be given priority in E-Learning. In this paper, our emphasis is laid on different risks faced by the participants and their remedies to make the system reliable and efficient.

The second section explains common threats, followed by different risks faced by the participants and necessity of risk analysis in E-Learning. Section three elaborates how these risks can be minimized by different tools and techniques. The fourth section offers a brief conclusion.

## 2. THREATS AND RISKS

A loss of an asset is caused by the realization of threats or risks. All threats /risks are realized through the medium of vulnerability. The major threats and as follows[7]:

a.  **Confidentiality violation :** An unauthorized party gaining access of the assets present in E-Learning system.
b.  **Integrity Violation :** An unauthorized party accessing and tempering with an asset used in E-Learning system.
c.  **Denial of Service :** Prevention of legitimate access rights by disrupting traffic during the transaction among the users of E-Learning system.
d.  **Illegitimate use :** Exploitation of privileges by legitimate users.
e.  **Malicious program :** Lines of code to damage the other programs.
f.  **Repudiation :** Persons denial of participation in any transaction of documents.
g.  **Masquerade:** A way of behaving that hides the truth by the hackers.
h.  **Traffic analysis:** Leakage of information by abusing communication channel.
i.  **Brute-force attack**: An attempt with all possible combinations to uncover the correct one.

As a result of above threats[2] following risks may occur during transaction of textual and non-textual messages among different participants of E-Learning system.

### 2.1. Author's risk

Modern technology has made it possible for Authors to provide access materials like books, journal papers, etc to a wide range of students, friends and acquaintances. Authors are responsible to develop and implement the contents. The reason why many Authors refrain from providing is the fear that their compiled material might be passed on and processed without the their knowledge. As only registered Students can access those lecturer notes, assignments, etc, it is the Author's duty to protect against unauthorized use, modification and reuse of the data in different contexts related to E-Learning.

Author's lecture notes, class test papers, home assignments etc. may be modified / destroyed by hackers through the above attacks. Therefore, it is in the Author's interest to ensure that the users receive the content unaltered and that the users can check the integrity of the text. Regular data backups and a plan of action in case of a breakdown of certain components (e.g. hard disk, network connections) are essential elements of a risk analysis. Financial interests also frequently play an important role in all such cases [11].

Only Authors know how much time is required for writing individual chapters and hence total books/materials. So it is the task of the group of Authors to contribute their viewpoints to  risk analysis.

## 2.2 Teacher's risk

Teachers are responsible for providing every possible support to the Students related to academic matter. Teachers may follow or buy the course content, presentations from a third party according to the requirement of the course.

All risks of E-Learning are not to be restricted to the technical system. It is necessary to cover the entire methods of teaching, examination, evaluation and grading. Teaching methodologies change from one Teacher to another but there will be common  risks in events such as delivering lecture, sending notes and assignments, accepting  and  marking answer sheets, preparing and distributing mark sheets.

Discussions are an essential component of teaching any course. One form of discussion can be through the online forum. An advantage of online forum discussions over oral discussions is that all written documents are stored electronically on a server, but the digital storage of contributions to a discussion constitutes a great risk for the privacy of Students as well as Teachers. Though in any teaching system maximum interaction can help Students as well as the Teachers to make their understanding clear .Only robust  security mechanism can lead to this kind of interaction in the long run [10].

There is a risk in the examination system which includes standardization of examination questions and list of questions possibly restrict the academic freedom of individual Teachers. Depending upon contract of jobs, a Teacher will play is role in academic centre. There must be a team to take care of all these risks.

Risk related to examination is directly associated with cheating. Apart from cheating, Teachers must be concerned about availability and non repudiation of assessments. Also at the time of examination Students are more eager to collect materials compared to studying content.

All Teachers must be aware of risk that  Students receive the unaltered questions paper before the beginning of the examinations and all answers are stored in an unaltered way as well. Though lecture is the most simple and natural form of communication,  there remains always a risk of modification of  the class lecture (speech) when it reaches  the Students.

## 2.3. Manager's risk

In any E-Learning system, the concern  Board or Authority grants diploma, degree or master certificate to a Student after successful completion of the course. This is like University Grant Commotion (UGC) or All India Council for Technical Education (AICTE), who is responsible for giving approval to any regular fulltime academic institute in India. But every Board always sets few rules and regulations for setting up and running  E-Learning institute. It becomes risky at the time of inspection if there is any ambiguity in following those rules.

Main risks in E-Learning involve inelegant people masquerading as Students and writing tests on behalf of enrolled Students and unauthorized help during the writing of online examination,

All faculty members and Students tend to neglect legal aspects because they are usually more interested in academic field. Legal issues such as copyright, online testing, sending official documents etc, may be a great risk for those participants. Managers should take care of enrolment in a course and the cancellation of enrolment as and when required. Enrolment of one particular student in more than one course involves risk for the larger organization. There must be a plan for backups and recovery process test .Otherwise at the time of requirement it will be difficult to make the data up to date.

It is also a risk for the management to distribute responsibilities in sensitive issues like maintaining password of all servers and routers, recordings of daily network traffic, looking after continuous power supply to the server and other network devices.

It is the duty of the Manager to control the authorisation i.e. access strategies (read, write and execute) to the Students and other participants for efficient running of the system. Other wise it may difficult to maintain privacy.

So Manager must assign system people (DBA) and authorise E-Learning system user (developer) to perform the operation like index i.e. allows creation and deletion of indices, Resources i.e. allows creation of new tables, Alteration allows addition or deletion of attributes in a relation, Drop i.e. allows deletion of relations .Also authorised to Read, Insert, Update, Delete on parts of multimedia databases of E-Learning materials. But the learner who has some form of authorisation may not allow to grant this authorisations to other learners and system administrator who has some authorisation may be allowed to withdraw (revoke) an authorisation that has granted earlier[8].

Besides all these, there are several other risks which should be looked after by the Manager. These are as follows

    i.      The server and individual PC may be affected by internet viruses, at the time of receiving e-mail or by running different software on that computer.

    ii.      Physical security of the building.

    iii.      Remote access, LAN, WAN damages.

    iv.      Training, Processing and document accessing problem.

    v.      LMS(learning management system) or CMS(content management system) damages

Also development costs can exceed initial estimates unless clear production goals are established and Implementation will be challenging if not well-planned in advance of development keeping in mind that not all content is suitable for delivery via E-Learning

## 2.4. System Developer's risk

In an existing system, there are some limiting factors. Some time, to improvise the entire system those factors have to be changed. But that is extensive and expensive also. In an E-Learning system, courses are classified into different modules. Due to the requirement of market demand one model (for example, the entire module of MCA has to be changed to MTech [Computer Science]) may be changed to another model. New development team will have to face different problem to maintain and implement new ones unless all modules have been designed earlier.

Designing, developing, and delivering E-Learning products requires a quality of hardware components such as high ended web server & database server ,high bandwidth internet leased line and a quality LMS, along with a robust infrastructure capable of sustaining multiple users and networked applications. System team should suggest the remedies of these risks properly otherwise total project cost will be almost double.

Another risk that developer must deal with storing passwords in clear text in the application code as an intelligent learner may be able to access the source code of the script and get access of the password of the databases. Also a password system may at risks or broken down when users' password can be stolen, changed by attackers. Today attacker are using many tools to guess the user's password (Ref Art 2.(i)).System developer or DBA must aware of SQL injection ,Cross-site scripting (XSS) attacks to maintain multimedia database.

## 2.5. Student's risk

Maximum number of users in E-Learning system is Students who learn as well as share their knowledge with other in the system. Student group can be classified into different levels from junior level, diploma, degree, post graduate, up to doctoral level. But every user must be aware of each and every material received from institute, Teachers or other Students. Other wise if intruders have edited the question papers or other important documents then the Students will have to face problems at the time of examination.

Risk of storing login information (user ID and passwords). All Students must be aware of misuse of login information, otherwise attacker [9] may attempt to prevent authorised learner from accessing the E-Learning server by the above attacks.

Teachers are not always available to help the Students so they need to be disciplined to work independently without the Teacher's assistance. Students also need to have good writing and communication skills. When Teachers and other Students aren't meeting face-to-face it is possible to misinterpret what was meant. As a feedback mechanism from Students will always enriches a Teacher, there is risk from Students side to send the same feedback to the management of the E-Learning institute.

At last all learners must be aware of phishing where attacker sets up fake web sites which look like a real E-Learning website so well   that human eye will not able to distinguish between real and attacker site. Here learners are prompted to enter some confidential information [3].

### 2.6. Others threats and risks in E-Learning

Besides the above risks there are various other threats present in the system as [4].

#### 2.6.1. Natural

Natural threats may be caused by natural disasters like fire, storm, volcanic eruption, earthquake, floods etc. E-Learning system can be affected highly by those threats.

#### 2.6.2. Deliberative act

 Threats may come from fraud, blackmail, theft etc.

#### 2.6.3. Unintended

There may be some unavoidable threats like Computer bug, power outage, handling error etc.

So all participants in E-Learning system must sit for a risk analysis where external IT and security experts could be included. Structuring of thoughts related to risks may be represented by different matrices.

## 3. REMEDIES OF RISKS

Participants of E-Learning system face different risks or threats as discussed in the previous section. Following tools or techniques may be imposed to minimize those risks.

### 3.1 Access control using Firewall

A firewall is a combination of hardware and software security system established to prevent unauthorized access to a corporate network from outside the organization. [15]Technically, a firewall is a specialized version of a router. Apart from the basic routing functions and rules, a router can be configured to perform the firewall functionality, with the help of additional software resources.

Main principle based on the rule is that all traffic from inside to outside and vice versa must pass through the firewall. To achieve this, all access to the local network must first be physically blocked, and access only via the firewall should be permitted. Only the traffic Authorized as per the local security policy should be allowed to pass through. The firewall itself must be strong enough, so as to render attacks on it useless.

In practical implementations, a firewall is usually a combination of packet filters and application (or circuit) gateways. One such firewall [3] is shown in Figure-1. So sophisticated firewalls can block some incoming traffic but permit E-Learning users (may be Students, Teacher, etc) to the inside to communicate freely from the outside.

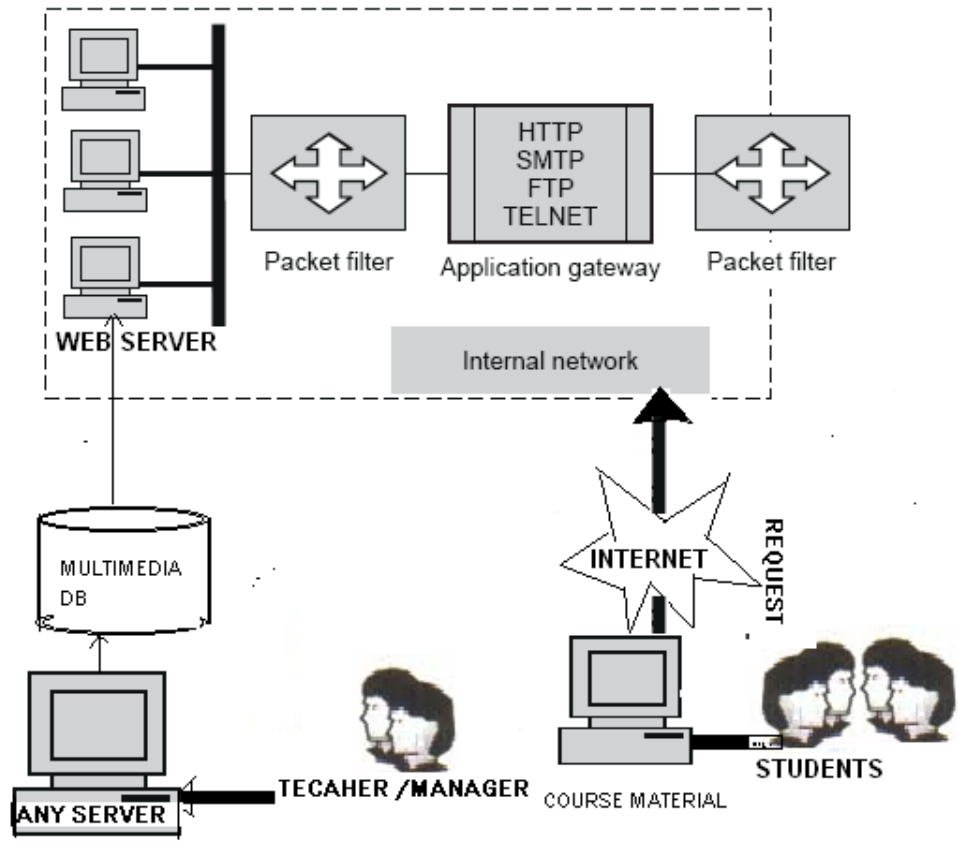

**Figure 1.  Organisation of  secure firewall based  in E-Learning system.**

So it is the duty of all system administrators to earn knowledge and skills to implement firewall, to configure the firewall and to monitor & troubleshoot firewalls.

### 3.2. Digital Right Management (DRM) on E-Learning assets

One of the major strategies to be implemented to reduce risks associated with E-Learning assets [12] is digital right management. Shareable asset is the simple resource, such as a static HTML page or a PDF document, or collection of files, such as images and a style-sheet. On the other hand asset of E-Learning system can be defined as E-Learning content (Exam, Notes, Grade), Cryptographic key content, User personal data, Messages between users, Different group membership data, Network bandwidth, Message integrity and Message availability. In this discussion, writers will define E-Learning asset as services provided by E-Learning system such as learning resources, examination or assessment questions, Students' results, user profile, forum contents, Students' assignment and announcement in the E-Learning system[12]. Digital Right Management (DRM) [1] makes the system safer for its contents. E-Learning system is working either in a distributed network or in Internet where multiple rights associated with learner, instructors content providers, administrators etc come into play as content and services are created, distributed, aggregated, disaggregated, stored found and used. That is why digitization is needed. In a general sense, DRM should be used for license agreement and copyright protection or prevents copying [16].

## 3.3. Cryptography

the purpose of confidentiality is to ensure that information and data are not disclosed to any unauthorized person or entity. Also readers must able to rely on the correctness of the course. One of the techniques in this aspect is cryptography [6]. Different cryptographic tools and techniques are needed for the implementation of security in Internet based transactions. There are two types of algorithms in cryptography

### 3.3.1. Secret-key algorithms

In secret-key algorithms the encryption & decryption key is the same, it requires the sender and receiver to agree on the key prior to the communication, the main function of this algorithm is encryption of data. Examples of such algorithms are Data Encryption Standard (DES), International Data Encryption Algorithms (IDEA), and Advanced Encryption Standard (AES).So only for encryption techniques for E-Learning content we can use these techniques.

### 3.3.2. Public-key algorithms

Public key cryptosystems, on the other hand, use one key (the public key) to encrypt messages or data, and a second key (the secret key) to decrypt those messages or data. Here three mathematical models are mainly used-Integer factorization, discrete logarithms and elliptic curve. Different public-key algorithms are RSA, El-Gamal, DiffieHellman.We can use these techniques  at the time of sending question paper and receiving answer sheets.
To authenticate a participant we can use following technologies using public key algorithm
  • Digital Signature
  • Digital certificate

## 3.4. Neural Cryptography

It is a new approach based on artificial neural networks (ANN) for data security in electronic communication. It is once again a cryptosystem, which is based on biological ideas including the network architecture, biological operations and the learning process. So the  complexity of the generation of the secured channel is linear with the size of the network. This biological mechanism may be used to construct an efficient encryption system using keys which change permanently[13]. It is very simple and fast to implement in context of possible attack at the time of transferring E-Learning document.

## 3.5. Elliptic Curve Cryptography (ECC)

As huge amount of textual and non-textual messages have to be transfer among participants so ECC will be  stronger option than any other cryptography techniques. It is tested  a popular key size require for RSA is 2,048 bits where   ECC requires 224 bits for same security. Also both confidentiality and authentication may be preserved in case of E-Learning document transfer using elliptic curve digital signature algorithm (ECDSA)[6].

## 3.6.  Biometric Authentication

Among all authentication techniques[5] like passwords, smart card, Digital signature and digital certificate, there is no guarantee that dishonest Students will keep their password secret. Password might be misused at the time of submission of assignment, receiving question papers, downloading of course materials, etc where biometric authenticity would give better security. But this needs a bit more capital investment.

### 3.7. Digital Watermarking

This technique allows an individual to add hidden copyright notices, audio, video, image signals. So multimedia database server of E-Learning system  may be protected against unauthorized use by the way of  digital watermarking. When also E-Learning  information like question papers ,important study materials ,etc will invisible to the viewer ,the chances of hacking will be nil or less.

## 4. CONCLUSIONS

We presented the risks that may occur by different participants of E-Learning and its counter measure tools/ techniques to minimize those risks. Though in E-Learning only the Student can unlock his private data ,rest all challenges remain on how to implement and maintain higher levels of privacy while setting up the learning process. Always the IT department strives to guarantee the availability of services  by using redundant hardware like server ,routers etc. Another important part that minimizes the risks is logs .Logs are distributed by virtue of the fact that they may be stored by different applications operating on different computers. Details of the transaction including the time of its occurrence would be "logged" and the resulting record will be  secured using cryptographic techniques. We can further improve the level of security in E-Learning by applying different other  techniques to minimize the risk though no system will be absolutely secured. Readers must be able to rely on the correctness of the content other wise by reading incorrect or non-relevant  content; readers will loose the  trust on the texts or will refuse to read for the next time onwards. In future , the concept of m-learning will come in new electronically learning features ,however new risks will also occur parallely with M-Learning.

**Authors**

**Mr. Nikhilesh Barik**

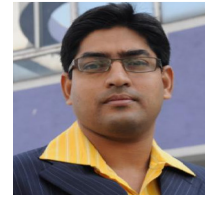

Mr. Nikhilesh Barik has completed B.Sc. Hons(Math) ,M.Sc(Math), MCA and M.Phil.(Computer Science) from different Indian universities. He was born on the 10th day of October in the year 1975. He is presently holding the post of Asst Professor in Computer Application Department at   Durgapur Society of Management Science, India since 2000. He is a research scholar under The University of Burdwan, India. His field of interest in research area is Network security.  Mr. Barik  has published / presented few papers related to security issues in e-learning, e-government and e-commerce.

**Dr. Sunil Karforma**

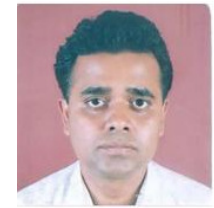

Dr Sunil Karforma has completed BE (Computer Science and Engineering) and ME (Computer Science and Engineering) from Jadavpur University. He has completed Ph. D. in the field of Cryptography. He is presently holding the post of Reader and the Head of the Department in the Department of Computer Science, The University of Burdwan. Network security and e-commerce are his field of interest in research area. He has published approximately 16 research papers in reputed National and International journals and proceedings.